\documentclass[journal=jacsat,manuscript=article]{achemso}

\usepackage[version=3]{mhchem} % Formula subscripts using \ce{}
\usepackage[T1]{fontenc}       % Use modern font encodings

\author{Sergi Lendinez}
\affiliation{Department of Physics and Astronomy, University of Delaware, Newark, DE 19716, USA}

\author{Mojtaba T. Kaffash}
\affiliation{Department of Physics and Astronomy, University of Delaware, Newark, DE 19716, USA}
\author{M. Benjamin Jungfleisch}
\affiliation{Department of Physics and Astronomy, University of Delaware, Newark, DE 19716, USA}
\email{mbj@udel.edu}

\title{Emergent spin dynamics enabled by lattice interactions in a bicomponent artificial spin ice}


\begin{document}

%%%%%%%%%%%%%%%%%%%%%%%%%%%%%%%%%%%%%%%%%%%%%%%%%%%%%%%%%%%%%%%%%%%%%
%% The "tocentry" environment can be used to create an entry for the
%% graphical table of contents. It is given here as some journals
%% require that it is printed as part of the abstract page. It will
%% be automatically moved as appropriate.
%%%%%%%%%%%%%%%%%%%%%%%%%%%%%%%%%%%%%%%%%%%%%%%%%%%%%%%%%%%%%%%%%%%%%
%\begin{tocentry}
%
%Some journals require a graphical entry for the Table of Contents.
%This should be laid out ``print ready'' so that the sizing of the
%text is correct.
%
%Inside the \texttt{tocentry} environment, the font used is Helvetica
%8\,pt, as required by \emph{Journal of the American Chemical
%Society}.
%
%The surrounding frame is 9\,cm by 3.5\,cm, which is the maximum
%permitted for  \emph{Journal of the American Chemical Society}
%graphical table of content entries. The box will not resize if the
%content is too big: instead it will overflow the edge of the box.
%
%This box and the associated title will always be printed on a
%separate page at the end of the document.
%
%\end{tocentry}

%%%%%%%%%%%%%%%%%%%%%%%%%%%%%%%%%%%%%%%%%%%%%%%%%%%%%%%%%%%%%%%%%%%%%
%% The abstract environment will automatically gobble the contents
%% if an abstract is not used by the target journal.
%%%%%%%%%%%%%%%%%%%%%%%%%%%%%%%%%%%%%%%%%%%%%%%%%%%%%%%%%%%%%%%%%%%%%
\begin{abstract}
 Artificial spin ice (ASI) are arrays on nanoscaled magnets that can serve both as models for frustration in atomic spin ice as well as for exploring new spin-wave-based strategies to transmit, process, and store information. Here, we exploit the intricate interplay of the magnetization dynamics of two dissimilar ferromagnetic metals arranged on complimentary lattice sites in a square ASI to effectively modulate the spin-wave properties. We show that the interaction between the two sublattices results in unique spectra attributed to each sublattice and we observe inter- and intra-lattice dynamics facilitated by the distinct  magnetization properties of the two materials. The dynamic properties are systematically studied by angular-dependent broadband ferromagnetic resonance and confirmed by micromagnetic simulations. We show that the combination of materials with dissimilar magnetic properties enables the realization of a wide range of two-dimensional structures potentially opening the door to new concepts in nano-magnonics.

\end{abstract}

%%%%%%%%%%%%%%%%%%%%%%%%%%%%%%%%%%%%%%%%%%%%%%%%%%%%%%%%%%%%%%%%%%%%%
%% Start the main part of the manuscript here.
%%%%%%%%%%%%%%%%%%%%%%%%%%%%%%%%%%%%%%%%%%%%%%%%%%%%%%%%%%%%%%%%%%%%%

Owing to their versatile properties, magnons, which are the elementary quanta of spin waves, have been explored in recent years as novel strategies to transmit, process, and store information in magnetic materials \cite{Kruglyak2010}. Magnons enable spin transport decoupled from electron motion, which potentially reduces the generation of waste heat \cite{Chumak_NatPhys_2015}. Their wavelength is highly tunable down to the technologically relevant sub-10 nm scale, whereas their resonance frequency can be adjusted from sub-one GHz to tens of THz. From a technological perspective, magnons have been envisioned as a platform for wave-based and neuromorphic computing \cite{Vogt2014,Bonetti2013,Grollier2016,Csaba_2017}, as well as a transduction mechanism for coherent coupling to different types of carriers such as photons and phonons \cite{Tabuchi_2014,Zhang_2016}.
To this end, the easily accessible nonlinear regime of spin waves and the ability to control magnon-magnon interactions are particularly important. In analogy to photonic crystals in photonics, magnonic crystals are artificial magnetic materials where the magnetic properties are periodically modulated in space allowing for an effective control of magnons and ultimately the manipulation of the spin-wave bandstructure \cite{Wang2009,Kostylev2008,Chumak_2017,Lenk_2011,Krawczyk_2014}.

In this context, nanoscopic artificial spin-ice (ASI) networks can be considered as a particular type of two-dimensional magnonic crystals \cite{Lendinez2020,Sklenar2019,Skjaervo_2019,Gliga2020,Dion_2019,Iacocca_2020}. Although ASI was originally introduced to model the frustrated behavior of atomic spin-ice systems \cite{Harris1997,Bramwell2001,Branford2012,Gilbert2016}, early computational and experimental work indicated that the reprogrammability of ASI facilitates the realization of novel functional magnonic materials \cite{Sklenar2013,Gliga2013,Jungfleisch2016,Zhou2016,Iacocca_PRB2016}.
In magnonic systems composed of nanowires, it was previously shown that the spin-wave band structure can be reconfigured by the proper choice of materials \cite{Gubbiotti2018}. However, this approach has not been applied to two dimensional structures such as ASI lattices where a wealth of possible configurations could potentially lead to unique characteristics including magnetic frustration and the capability to fine-tune the interaction between the nanoscaled elements \cite{Ostman2018,Drisko2017,Schanilec2019,Farhan2019,Gliga2020}. 

In this work, we introduce an artificial spin-ice network based on nanomagnets made of two dissimilar ferromagnetic metals (Ni$_{81}$Fe$_{19}$ and Co$_{90}$Fe$_{10}$) arranged on complimentary lattice sites in a square lattice. We show that ferromagnetic resonance spectroscopy is an effective tool to probe the interactions in the network. Using an angular-dependent ferromagnetic resonance approach, we discover unique spectra attributed to each sublattice, and -- even more importantly -- the observation of intra- and inter-lattice dynamics facilitated by the distinct magnetization properties of the two materials. The dynamics of the whole magnonic system are controlled by the interactions between the sublattices, and the choice of materials can be used to finely tune the spin-wave resonances. Time-dependent micromagnetic simulations \cite{Vansteenkiste2014} reveal the detailed excitation profiles in the sublattices, further demonstrating the potential to modulate the exact spin-wave modes in the nanoscale network.
Combining materials with different magnetic properties in the artificial spin ice enables the realization of a wide range of two-dimensional structures opening the door to new concepts in nano-magnonics

\begin{figure*}[t]
    \centering
    \includegraphics{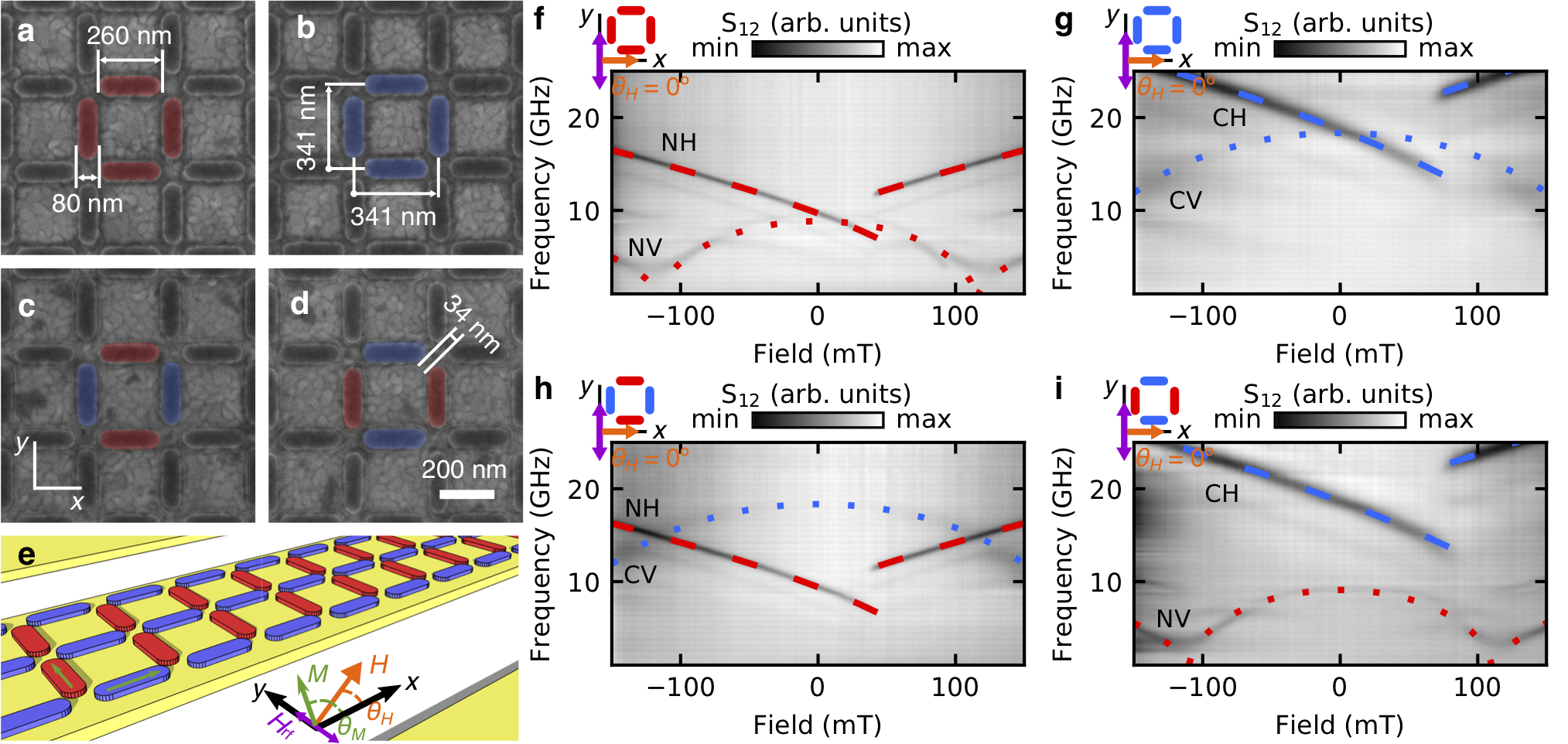}
    \caption{\textbf{Characterization of single- and bicomponent artificial spin ices.} Scanning electron microscopy images of studied square lattices and their dimensions for \textbf{a} single-component Ni$_{81}$Fe$_{19}$, \textbf{b} single-component Co$_{90}$Fe$_{10}$, \textbf{c} bicomponent Ni$_{81}$Fe$_{19}$-Co$_{90}$Fe$_{10}$ with Ni$_{81}$Fe$_{19}$ in the horizontal sublattice, and \textbf{d} bicomponent Ni$_{81}$Fe$_{19}$-Co$_{90}$Fe$_{10}$ with the Co$_{90}$Fe$_{10}$ in the horizontal sublattice. The red false color represents nanomagnets made of Ni$_{81}$Fe$_{19}$, and the blue false color represents nanomagnets made of Co$_{90}$Fe$_{10}$. In \textbf{e}, a schematic of a coplanar waveguide with a bicomponent spin ice on the signal line is shown. The corresponding absorption spectra at magnetic fields applied parallel to the signal line are shown in \textbf{f}--\textbf{i}. Horizontal elements are fitted to equation.~\eqref{eq:fmr_ea}: Ni$_{81}$Fe$_{19}$ modes, NH (red dash curves); Co$_{90}$Fe$_{10}$ modes, CH (blue dash curves). Vertical elements are fitted to equation~\eqref{eq:fmr_ea} at high fields, and using Supplementary equation~(S1.2) at low fields: Ni$_{81}$Fe$_{19}$ modes, NV (red dot curves); Co$_{90}$Fe$_{10}$ modes, CV (blue dot curves).}
    \label{fig:setup}
\end{figure*}

\section*{Results}
%\subsection*{Resonant dynamics}
%\subsection*{Fabrication \textcolor{blue}{and experimental setup}}
%We used a multi-step lithography process for sample fabrication (see Methods for details). %Different types of square ASI were fabricated by electron-beam lithography on top of the signal line of coplanar waveguides (CPWs) to measure the dynamic response under microwave excitation. 
%The nanomagnets were designed to have a length of 260 nm, width of 80 nm, and thickness of 20 nm (as indicated in Fig.~\ref{fig:setup}a), and arranged in square lattices with unit vectors of 341 nm (as indicated in Fig.~\ref{fig:setup}b) in $x$ and $y$ directions (coordinate system shown in Fig.~\ref{fig:setup}c), corresponding to a minimum gap of 34 nm (as indicated in Fig.~\ref{fig:setup}d). The same geometry and dimensions, but different materials, were used for four different samples shown in Fig.~\ref{fig:setup}:
We fabricated a set of ASI arranged in a nominally identical square geometry, but with different material combinations (see Supporting Information for details): (A) All elements consist of Ni$_{81}$Fe$_{19}$ (Fig.~\ref{fig:setup}a), (B) all elements consist of Co$_{90}$Fe$_{10}$ (Fig.~\ref{fig:setup}b), (C) horizontal elements are made of Ni$_{81}$Fe$_{19}$ while vertical elements are made of Co$_{90}$Fe$_{10}$ (Fig.~\ref{fig:setup}c), and (D) horizontal elements are made of Co$_{90}$Fe$_{10}$, while vertical elements are made of Ni$_{81}$Fe$_{19}$ (Fig.~\ref{fig:setup}d). %The thickness of the materials was chosen so that a single-domain state in each element is facilitated. Scanning electron microscope (SEM) images of the different samples are shown in Figs.~\ref{fig:setup}A-D, where false colored elements are super-imposed in one extended unit cell to highlight the different material constituents in each sample (red color represents Ni$_{81}$Fe$_{19}$ and blue color represents Co$_{90}$Fe$_{10}$.) 
A schematic drawing of the measurement setup is shown in Fig.~\ref{fig:setup}e. We use a standard vector network analyzer (VNA) ferromagnetic resonance (FMR) technique \cite{Kalarickal2006}, see Supporting Information.

%, in which a high-frequency signal is sent through the CPW (yellow lines depicted in Fig.~\ref{fig:setup}E) creating an oscillatory microwave magnetic field $H_\mathrm{rf}$ perpendicular to the signal line (in $y$ direction, purple arrow in the coordinate axis). In resonance, the microwave magnetic field exerts a torque on the magnetic moments in the magnetic elements patterned on top of the signal line. We measure the transmitted signal (S$_{12}$) using the VNA as a function of the magnetic field. In this way, we obtain color-coded images in which a decreased signal is observed when the microwave magnetic field efficiently excites dynamics in the ASI. % and hence the transmitted power is minimum (darker colors in Figs.~\ref{fig:setup}F-I) and the absorbed power is maximum.

%\subsection*{Resonant dynamics}
Figures~\ref{fig:setup}f-i show the experimentally detected absorption spectra with an applied magnetic field parallel to the coplanar waveguide (CPW) signal line ($\theta_H=0^\circ$) as false color coded images, where a dark contrast represents maximum absorption (minimum in the transmission parameter S$_{12}$), while a light contrast represents minimum absorption (maximum in the transmission parameter S$_{12}$). %\sout{Note that the horizontal, field-independent lines in the spectra are an artifact caused by the normalization of the spectra.} 
The applied magnetic field in all presented measurements is swept from negative to positive field values.
%Before comparing the experimental FMR spectra with systematic micromagnetic simulations, we discuss their general behaviors as well as the origin of the most pronounced modes.
%A glaring difference between the lattices becomes evident just by looking at the spectra shown in Figs.~\ref{fig:setup}F-I without inspecting the detailed mode structure. 
We can identify the main absorption lines corresponding to the different lattice sites by correlating the observed spectra with the sample configurations taking into account the magnetic properties of Ni$_{81}$Fe$_{19}$ and Co$_{90}$Fe$_{10}$. 

Following the general expression given by Suhl \cite{Suhl1955}, the resonance frequency $f$ in an in-plane geometry is given by (see Supplementary Fig. S1 for the symbol definition):
%\begin{widetext}
%\begin{equation}
%\textcolor{blue}{f = %\frac{\gamma}{2\pi}\mu_0\sqrt{[H_\mathrm{ext}+(N_\mathrm{z}-N_\mathrm{\parallel}) M_\mathrm{s}]\times [H_\mathrm{ext}+(N_\perp - N_{\parallel}) M_\mathrm{s}]}},
%\label{Eq:Kittel}
%\end{equation}
%\end{widetext}
%where $f$ is the resonance frequency, $\gamma$ is the gryomagnetic ratio, $H_\mathrm{ext}$ is the applied magnetic field, $N_\mathrm{z}$ is the demagnetizing factor along the film thickness ($z$-direction), and $N_\parallel$, $N_\perp$ are the demagnetizing factors of the elements parallel and perpendicular to $H_\mathrm{ext}$, respectively, and $M_\mathrm{s}$ is the saturation magnetization. Ni$_{81}$Fe$_{19}$ is magnetically softer than Co$_{90}$Fe$_{10}$ ($M_\mathrm{s}(\mathrm{NiFe}) = 800 \times 10^3$~A/m, $M_\mathrm{s}(\mathrm{FeCo}) = 1350 \times 10^3$~A/m), and thus according to Eq.~(\ref{Eq:Kittel}) the resonance frequency of Ni$_{81}$Fe$_{19}$ at a given field value is lower than that of the Co$_{90}$Fe$_{10}$ elements.}
%\sout{For the horizontal elements, where the magnetic field $\mu_0 H$ is applied along the easy axis, the resonance frequency is given by:}
%\begin{widetext}

\begin{align}
%\textcolor{magenta}{f = \frac{\gamma}{2\pi}\mu_0\sqrt{( H+H_1) ( H+H_2)}},
f & =  \frac{\gamma}{2\pi}\mu_0\sqrt{\left[H \cos(\theta_H-\theta_M)+M_\mathrm{S} (N_y - N_x) \cos (2 \theta_M)\right]} \nonumber\\
 & \quad  \cdot \sqrt{ \left[H \cos (\theta_H-\theta_M)+M_\mathrm{S} \left(N_z-N_x \cos ^2(\theta_M)-N_y \sin ^2(\theta_M)\right)\right]}
\label{eq:fmr}
\end{align}

%\end{widetext}
where $\gamma$ is the gyromagnetic ratio, $\mu_0$ is the magnetic permeability, $H$ is the external magnetic field, $M_\textrm{S}$ the saturation magnetization, $\theta_{H,M}$ the angle between the $x$ axis and the external magnetic field and the magnetization, respectively, and $N_{x,y,z}$ the demagnetization factors along each axis. In the derivation of equation~\eqref{eq:fmr}, we have considered a negligible magnetocrystalline anisotropy (confirmed in thin film data, shown in Supplementary Fig.~S2). 

In the case of a magnetic field applied along the $x$ axis, $\theta_H=0$, and the magnetization pointing along the magnetic field ($\theta_M=\theta_H=0$), the expression can be reduced to

\begin{equation}
    f=\frac{\gamma}{2\pi}\mu_0\sqrt{[H+(N_y-N_x)M_\mathrm{S}][H+(N_z-N_x)M_\mathrm{S}]},
    \label{eq:fmr_ea}
\end{equation}

similar to previous results \cite{Zhou2016}.

For the horizontal elements, the magnetic field is applied along the easy axis, and equation~\eqref{eq:fmr_ea} can be used with $N_y>N_z>N_x$.

For the vertical elements, the value of $\theta_M$ is not constant at low fields. For instance, due to the shape anisotropy, the magnetization points perpendicular to the $x$ axis ($\theta_M=\pi/2$) at zero field, and tilts towards the $x$ axis as the magnetic field gradually increases. When the field is strong enough, the magnetization is aligned with the external field and the $x$ axis ($\theta_M=0$). Hence, $\cos(\theta_M)=H/H_1$, where $H_1\sim M_\textrm{S}$ is the field at which the magnetization is saturated. Once this field is reached, equation~\eqref{eq:fmr_ea} applies. Due to the shape anisotropy in the vertical elements, $N_x>N_y$ and $N_x>N_z$, the resulting resonant frequency is lower than in the horizontal elements, where $N_y>N_x$ (see Supporting Information for all expressions).

%For the vertical elements, the assumptions to derive Eq.~\eqref{eq:fmr_ea} are no longer valid, namely, that the magnetization points in the direction of the external field, and the resonant frequency is given by: \textcolor{red}{\textbf{Reference}}
%\begin{equation}
%\textcolor{magenta}{f=}
%\begin{cases}
%\textcolor{magenta}{\frac{\gamma}{2\pi}\mu_0\sqrt{(H_1 H_2-H^2\cdot\frac{H_2}{H_1})}} & \textcolor{magenta}{\textrm{for }H<H_1}\\
%\textcolor{magenta}{\frac{\gamma}{2\pi}\mu_0\sqrt{(H+(H_2-H_1))(H-H_1)}} & \textcolor{magenta}{\textrm{for }H\geq H_1}
%    
%\end{cases}
%,
%\label{eq:fmr_ha}
%\end{equation}
%\textcolor{magenta}{
%where we have taken into account that, for magnetic fields below the effective anisotropy field $H_1$, the magnetization deviates from the direction of the external magnetic field ($\textrm{arccos}(\theta_{M-H})=H/H_1$, with $\theta_{M-H}$ the angle between the magnetization $M$ and the external field $H$).
%}

We first inspect the single-component Ni$_{81}$Fe$_{19}$ network: Figure~\ref{fig:setup}f shows the corresponding spectrum where two main modes originating from the vertical and the horizontal elements are observed. When the sample is saturated, the magnetic moments in the horizontal elements lie in the easy axis for $\theta_M=\theta_H=0^\circ$ and equation~\eqref{eq:fmr_ea} applies with $N_y>N_x$ and $N_z\approx 0$. The horizontal Ni$_{81}$Fe$_{19}$ absorption lines start at 16.5~GHz for $-150$~mT, and decrease as the field is ramped up (labeled as NH in Fig.~\ref{fig:setup}f). When a field $\mu_0 H_\mathrm{NiFe}=50$~mT is reached, the magnetization of the horizontal elements switches, producing a step in the absorption frequency and a change in the sign of the slope. A fit according to equation~\eqref{eq:fmr_ea} is shown as a red dashed line in Fig.~\ref{fig:setup}f. For the vertical elements, the magnetic field is applied along the hard axis. The absorption lines start at 4.8~GHz for $-150$~mT (labeled as NV in Fig.~\ref{fig:setup}f) and decrease down to 3.4~GHz at a field of $-123$~mT, where the magnetization configuration in the vertical elements becomes unstable and then gradually rotates from the hard ($\theta_M=180^\circ$) to the easy axis direction ($\theta_M=90^\circ$). As the field decreases and the magnetization rotates to the easy axis, the resonance frequency increases until zero field is reached {producing a characteristic W-shape \cite{Montoncello_2018,Bang_2020}, which is a convenient way to qualitatively inspect the FMR spectra.} This mode is symmetric around zero field and the same behavior is found at positive fields.  It is important to point out that the torque exerted by the microwave magnetic field $H_\mathrm{rf}$ on the vertical elements becomes minimum when the magnetization lies in the easy axis parallel to $H_\mathrm{rf}$, which results in a gradually decreasing absorption. The red dotted curves in Fig.~\ref{fig:setup}f and i have been obtained by fitting the NV mode taking into account the different behavior at high and low fields, see also Supplementary equation~(S1.2).
% Note that, experimentally, the frequency at the effective anisotropy field $H_1^{NV}$ is not 0 but instead has a smooth minimum, unlike the curves from Eq.~\eqref{eq:fmr_ha}. This is produced by the missalignment of parts of the magnetization, especially around the edges, deviating from the macrospin picture used to derive Eq.~\eqref{eq:fmr_ha}.} % \textcolor{blue}{ Additional modes visible in the spectrum will be discussed below in comparison to systematic micromagnetic simulations.}

The same general behavior is observed for the single-component Co$_{90}$Fe$_{10}$ ASI (sample B) shown in Fig.~\ref{fig:setup}g. However, the resonance frequencies are pushed up in comparison to Ni$_{81}$Fe$_{19}$ due to the higher $M_\mathrm{S}$. In this case, the mode from the horizontal islands is labeled CH and the mode from the vertical islands CV, and the fits to the analytical expressions are shown in blue dashed and blue dotted curves, respectively. In addition, the absorption lines are noticeably broader due to the larger Gilbert damping of Co$_{90}$Fe$_{10}$ as revealed by the thin-film data (shown in Supplementary Fig.~S2). %\textcolor{blue}{The main mode coming from the horizontal elements (labeled CH in Fig.~\ref{fig:setup}g), is fitted using Eq.~\eqref{eq:fmr_ea} and shown as blue dashed lines. }%The intense mode that starts around 25~GHz at $-150$~mT, and decreases in frequency as the field is reduced, is caused by the horizontal elements (labeled CH in Fig.~\ref{fig:setup}g), \textcolor{blue}{is fitted using Eq.~\eqref{eq:fmr_ea} and shon as blue dashed lines}. %At a field of $\mu_0 H_\mathrm{CoFe}=70$~mT the Co$_{90}$Fe$_{10}$ magnetization abruptly switches producing a step in the spectrum. Again, the switching field for Co$_{90}$Fe$_{10}$ is higher than that for Ni$_{81}$Fe$_{19}$ due to the higher value of $M_\mathrm{S}$. The fitting to the CH mode using Eq.~\eqref{eq:fmr_ea} is shown in Fig.~\ref{fig:setup}G with blue dashed lines.
%The modes from the vertical Co$_{90}$Fe$_{10}$ islands are only visible as a faint line that starts slightly above 10~GHz at $-150$~mT and increases in frequency as the magnetic field approaches 0~mT (labeled CV in Fig.~\ref{fig:setup}g). Within the measured range, the external magnetic field is not strong enough to oppose the shape anisotropy and to saturate the magnetization along the hard axis. The dotted blue line in Fig.~\ref{fig:setup}g shows the fitting to the vertical elements in the low-field regime, where the magnetization is rotating toward the easy axis as the magnetic field is reduced. %This mode disappears between $-100$ mT and 100 mT as the torque generated by the CPW becomes minimum. This is analogous to the low-field behavior of the corresponding mode in the single-component Ni$_{81}$Fe$_{19}$ lattice in Fig.~~\ref{fig:setup}F.

The FMR spectra observed for the two bicomponent lattices are shown in Figs.~\ref{fig:setup}h and i. The spectra can be understood as a superposition of the mode structure caused by either of the two components Co$_{90}$Fe$_{10}$, Ni$_{81}$Fe$_{19}$ and the sublattice site they are occupying. For sample C (Fig.~\ref{fig:setup}h) the NH mode starting at 16~GHz at $-150$~mT corresponds to the horizontal Ni$_{81}$Fe$_{19}$ elements, similar to sample A as shown in Fig.~\ref{fig:setup}f, while the NV mode corresponding to the vertical Ni$_{81}$Fe$_{19}$ elements is absent. Instead, we observe the CV mode of the vertical Co$_{90}$Fe$_{10}$ islands, similar to sample B as shown in Fig.~\ref{fig:setup}g. When the two materials on the different sublattice sites are exchanged, the complimentary superposition of modes is observed; see Fig.~\ref{fig:setup}i.

In the following, we demonstrate the tunability of the resonant dynamics by modifying the lattice interactions. We rotate the in-plane magnetic field direction with respect to the CPW guide axis, $\theta_H$, which alters the magnetization configuration in the islands, and hence the interaction among them. This tuning mechanism could also be used for magnon-magnon coupling and hybridized magnonic states \cite{Ono, Li_2020, Liu_2019}. Besides, we will show that the altered magnetization configuration enables the enhancement of the detection sensitivity to the vertical islands. 

\begin{figure}[t]
    \centering
    \includegraphics{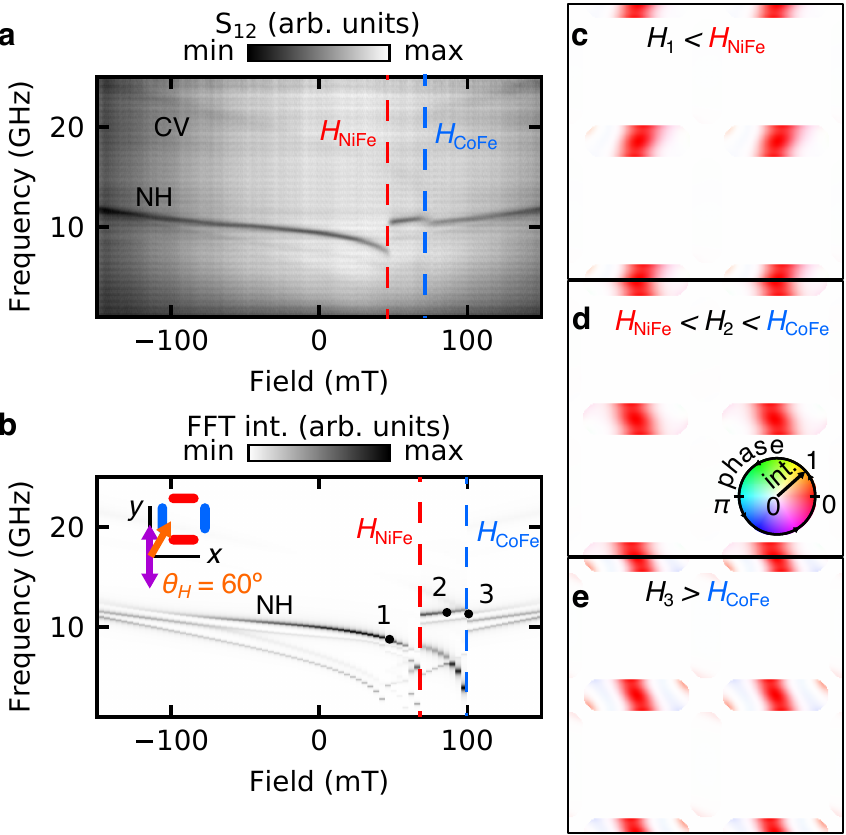}
    \caption{\textbf{Results for the bicomponent sample with the Ni$_{81}$Fe$_{19}$ sublattice in the direction of the signal line with external fields applied at 60 degrees.} The measured spectra are shown in \textbf{a}, which are in agreement with the spectra obtained in micromagnetic simulations shown in \textbf{b}. The profiles of the main resonant mode{, NH,} are shown {at three representative} magnetic fields {marked in \textbf{b}}: {1)} Before the Ni$_{81}$Fe$_{19}$ switching {(50 mT, 8.6 GHz)} in \textbf{c}{; 2)} after the Ni$_{81}$Fe$_{19}$ switching and before the Co$_{90}$Fe$_{10}$ switching {(80 mT, 11.6 GHz)} in \textbf{d}; and {3)} after the Co$_{90}$Fe$_{10}$ switching {(100 mT, 11.5 GHz)} in \textbf{e}. {The profile plots depict the FFT intensity with the saturation (white being zero intensity, and high saturation being maximum intensity normalized to 1) and the phase as the hue, as shown in the \textit{hsv} color wheel.}}
    \label{fig:60deg}
\end{figure}

Figure~\ref{fig:60deg} illustrates the results for sample C (Ni$_{81}$Fe$_{19}$ on the horizontal islands and Co$_{90}$Fe$_{10}$ on the vertical islands) with $\theta_H=60^\circ$. % with respect to the $x$ axis. 
As is obvious from the experimental results, Fig.~\ref{fig:60deg}a, the main NH mode shifts down when $\theta_H$ is increased to $60^\circ$ (it starts at 12~GHz at $-150$~mT). With increasing field, the magnetization of the Ni$_{81}$Fe$_{19}$ islands switch at $\mu_0 H_\textrm{NiFe}=47$~mT, resulting in a step-like increase in frequency and a change in sign of the slope \cite{Montoncello_2018,Bang_2020}. When the field is further increased to $\mu_0 H_\textrm{CoFe}=74$~mT, the vertical Co$_{90}$Fe$_{10}$ islands switch and the NH mode shows a step-like decrease in frequency. The switching of the Co$_{90}$Fe$_{10}$ islands can also be seen in the the faint CV mode at higher frequencies. Thus, the utilization of a material with different switching field leads to emergent spin dynamics absent in single-component ASI. This method can be used to tune the dynamic response of the sublattice, making the dynamic behavior different from just the addition of the two sublattices (which manifests in the superposition of the spectra).

%The experimental results for sample C at applied magnetic fields of $\theta=60^\circ$ with respect to the CPW signal line show an additional step in the main mode NH, Fig.~\ref{fig:60deg}(a). Besides the step increase and change in slope in the main mode frequency of sample C at $\mu_0 H_\mathrm{NiFe}=50$~mT when the NiFe islands switch, there is an additional step to smaller frequencies at $\mu_0 H_\mathrm{CoFe}=75$~mT. This field corresponds to the switching of the vertical CoFe islands, as can be seen from the high frequency line corresponding to the CoFe mode CV. Note, then, how the choice of a different material with different switching fields can be used to tune the dynamic response of the sublattice\textcolor{magenta}{, making the dynamic behavior different from just the addition of the two sublattices.}

The results of the corresponding simulations qualitatively agree with the experimental data, as shown in Fig.~\ref{fig:60deg}b. The trend of the main mode NH is reproduced, as well as the steps in frequency when the different sublattices switch; first the Ni$_{81}$Fe$_{19}$ sublattice (indicated by vertical red dashed lines in Figs.~\ref{fig:60deg}a and b), and then the Co$_{90}$Fe$_{10}$ sublattice (indicated by vertical blue dashed lines in Figs.~\ref{fig:60deg}a and b). Note that the step at $\mu_0 H_\mathrm{CoFe}$ is not observed in sample A, composed of only Ni$_{81}$Fe$_{19}$, as shown in Supplementary Fig.~S3. %\textcolor{red}{Would it make sense to refer to 1, 2, 3 in the figure? PROBABLY NOT HERE, BUT FARTHER DOWN -- see below}. 
The values of the resonant frequency and the switching fields do not exactly match the ones observed in the experiment, which may be caused by slightly different values of parameters used in the simulations such as saturation magnetization, geometry, and exact angle of the applied magnetic field. %Despite these minor differences, we emphasize that the qualitative behavior is reproduced in the micromagnetic modeling, which allows us to further confirm that the steps in the absorption lines are due to the reversal of the different sublattices and to gain insights into the spatial dynamic profiles.

In addition to the absorption spectra, we determine the excitation profiles of the lattice structures by micromagnetic simulations. In this way, we can characterize the different modes and determine the spatial location of particular excitations. Besides the spatial profile of the FFT intensity, we also obtain the phase of the oscillations, and map both the intensity and the phase using the \textit{hsv} color space, where \textit{h} stands for hue, \textit{s} for saturation, and \textit{v} for value. Here, the phase is represented by the hue and the intensity by the saturation, with the value being fixed to 1. We find that the most intense curve in Fig.~\ref{fig:60deg}b (labeled as NH) corresponds to oscillations coming from the center of the horizontal Ni$_{81}$Fe$_{19}$ islands. Figure~\ref{fig:60deg}c shows the excitation profile at point 1 in the simulation shown in Fig.~\ref{fig:60deg}b before the Ni$_{81}$Fe$_{19}$ islands switch. This first-order oscillation mode is characterized by an anti-node with maximum intensity in the center of the horizontal island and two nodes in the edges of the island. As the Ni$_{81}$Fe$_{19}$ islands reverse their magnetization at $\mu_0 H_\mathrm{NiFe}$, the oscillation remains localized in the center of the Ni$_{81}$Fe$_{19}$ islands with the nodes staying at the edges, and a slightly opposite tilt determined by the relative orientation of the magnetization and the external field. This is depicted in Fig.~\ref{fig:60deg}d, corresponding to point 2 in Fig.~\ref{fig:60deg}b. When the vertical Co$_{90}$Fe$_{10}$ islands switch at $\mu_0 H_\mathrm{CoFe}$, the frequency drops and a higher order excitation in the Ni$_{81}$Fe$_{19}$ islands replaces the fundamental mode. In this higher-order mode, the edges of the horizontal islands oscillate and two additional nodes appear between the center and the edges at each side. Between the nodes, a small, less-intense oscillation occurs out of phase, as shown in Fig.~\ref{fig:60deg}e (corresponding to point 3 in Fig.~\ref{fig:60deg}b).

\begin{figure}[t]
    \centering
    \includegraphics{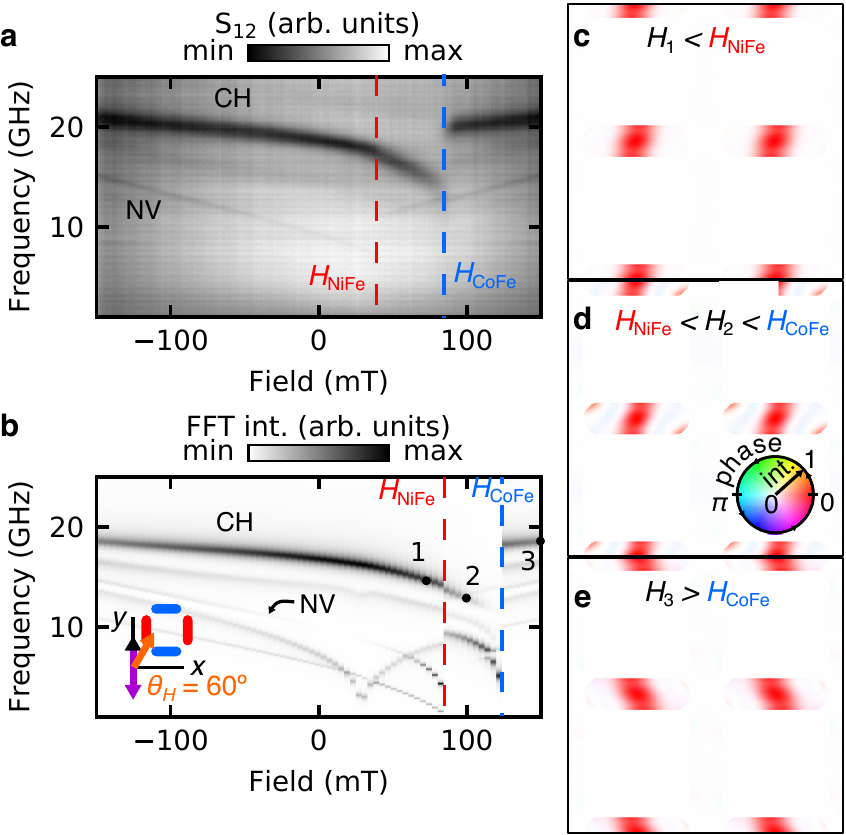}
    \caption{\textbf{Results for the bicomponent sample with the Co$_{90}$Fe$_{10}$ sublattice in the direction of the signal line with external fields applied at 60 degrees.} The measured spectra are shown in \textbf{a}, which are in agreement with the spectra obtained in micromagnetic simulations shown in \textbf{b}. The profiles of the main resonant mode{, CH,} are shown {at three representative} magnetic fields {marked in \textbf{b}}: {1)} Before the nd the Ni$_{81}$Fe$_{19}$ switching {(80 mT, 14.4 GHz)} in \textbf{c}{; 2)} after the Ni$_{81}$Fe$_{19}$ switching and before the Co$_{90}$Fe$_{10}$ switching {(100 mT, 12.8 GHz)} in \textbf{d}; and {3)} after the Co$_{90}$Fe$_{10}$ switching {(150 mT, 18.6 GHz)} in \textbf{e}. {The profile plots depict the FFT intensity with the saturation (white being zero intensity, and high saturation being maximum intensity normalized to 1) and the phase as the hue, as shown in the \textit{hsv} color wheel.}}
    \label{fig:60deg-comp}
\end{figure}

Conversely, a similar behavior can be observed in sample D, where the Co$_{90}$Fe$_{10}$ and Ni$_{81}$Fe$_{19}$ islands are exchanged, Fig.~\ref{fig:60deg-comp}. 
In the experimental data (Fig.~\ref{fig:60deg-comp}a) the Co$_{90}$Fe$_{10}$ main mode (labeled CH) from the horizontal islands shows a small step to lower frequencies at a magnetic field of $\mu_0 H_\mathrm{NiFe} = 39$~mT, when the vertical Ni$_{81}$Fe$_{19}$ islands switch. The NV mode from the vertical islands (faintly visible at smaller frequencies) also exhibits a step-like increase and change in sign of the slope, corroborating that it is indeed the vertical Ni$_{81}$Fe$_{19}$ islands reversing. Additionally to these indications, the small step in the CH mode is absent in the data from sample B (single-component  Co$_{90}$Fe$_{10}$), see Supplementary Fig.~S4. At $\mu_0 H_\mathrm{CoFe}=83$~mT the horizontal Co$_{90}$Fe$_{10}$ islands reverse, resulting in a step-like increase in the CH main mode frequency and a change in sign of the slope. At the same time, we observe a small step-like decrease in the NV mode frequency. The same qualitative trends are obtained in the micromagnetic simulations shown in Fig.~\ref{fig:60deg-comp}b for the CH and NV modes. The micromagnetic simulations show additional modes not obersved in the experiments, produced by higher order modes and modes localized at the edges of the samples. While these modes might be excited in our experiments, their localized nature at the ends of the islands leads to an intensity that is too small to be  detected experimentally.

Similarly to the case of sample C, the most intense CH mode is due to the central regions of the horizontal islands (Fig.~\ref{fig:60deg-comp}c shows the excitation profile at point 1 in Fig.~\ref{fig:60deg-comp}b), before the Ni$_{81}$Fe$_{19}$ islands switch. %Here, the mode lies at higher frequencies due to the higher saturation magnetization of Co$_{90}$Fe$_{10}$. 
When the Ni$_{81}$Fe$_{19}$  vertical islands switch, the frequency drops and a higher order excitation occurs, as shown in Fig.~\ref{fig:60deg-comp}d, corresponding to the excitation profile at point 2 in Fig.~\ref{fig:60deg-comp}b. This is analogous to the switching of the Co$_{90}$Fe$_{10}$ vertical islands in sample C. Finally, as the Co$_{90}$Fe$_{10}$ horizontal islands of sample D switch, the central oscillation region tilts slightly in the opposite direction as the magnetization aligns with the external field, as shown in Fig.~\ref{fig:60deg-comp}e corresponding to the excitation profile at point 3 in Fig.~\ref{fig:60deg-comp}b. The change in the tilt is in agreement with the behavior of sample C when the Ni$_{81}$Fe$_{19}$ horizontal islands switch.

\begin{figure}
    \centering
    \includegraphics{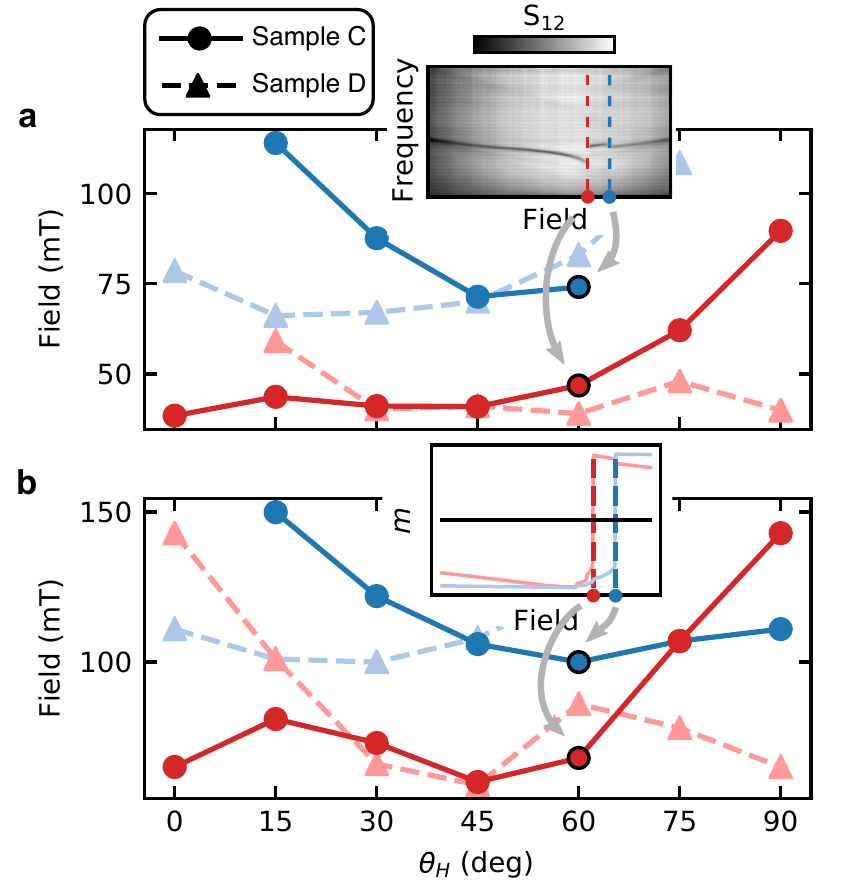}
    \caption{\textbf{Angular dependence of the switching fields in the bicomponent ASIs.} Change in the switching fields for sample C (dark solid lines and dots) and sample D (light dashed lines and triangles), for both the Ni$_{81}$Fe$_{19}$ sublattice (red curves) and the Co$_{90}$Fe$_{10}$ sublattice (blue curves), obtained in experiments \textbf{a} and simulations \textbf{b}. The switching fields from the experimental data in \textbf{a} are obtained from the dynamic measurements (as shown in the inset), and the switching fields from the simulated data in \textbf{b} are obtained from the simulated hysteresis loops (as shown in the inset).}
    \label{fig:switch}
\end{figure}

Although the exact field values are different, a similar trend is observed in the switching field values obtained from experiment and simulation, as summarized in Fig.~\ref{fig:switch} as a function of the in-plane field angel $\theta_{H}$. Interchanging the sublattices from sample C to sample D results in the expected symmetric behavior around $\theta_H=45^\circ$ of the switching fields, as is apparent from the comparison between the curves for samples C and D shown in Fig.~\ref{fig:switch}a. The choice of particular materials and their position on the specific lattice sites affect the intricate dynamics of the ASI. For instance, the micromagnetic simulations (Fig.~\ref{fig:switch}b) show that it is possible to change which sublattice switches at lower fields by just varying the magnetic field angle. Our results show that the interaction and the dynamics can be modified by a proper choice of the materials. In contrast, in single-element lattices, the switching of both sublattices occurs at  similar field values (see Supplementary Fig.~S5). Thus, by using dissimilar materials, we are able to extend the region at which only one sublattice has switched while the complementary sublattice retains the original magnetization configuration, allowing for an extended region in which intermediate modes can exist.

%\section*{\label{sec:conclusions}Conclusions}

In summary, we showed that using a bicomponent artificial spin ice arranged on a square lattice, we can finely tune the interaction between the lattice sites. We demonstrate that ferromagnetic resonance is an effective tool to probe this interaction. The resonant dynamics in the nanoscale network can be precisely adjusted by changing the magnitude and the angle of the applied magnetic field.  As the islands of one sublattice switch, the dynamics of the opposite sublattice is affected due to the change in the local dipolar field distribution. Thus, the choice of materials determines the particular switching field of each sublattice, and a further adjustment can be achieved by changing the external magnetic field direction. The reported tuning mechanism of the inter- and intra-lattice interactions offers a richer manipulation of the specific mode spectra than the previously reported \textit{superposition principle} of the resonant dynamics in single-component artificial spin ice. 
Our results demonstrate the ability to realize a novel type of two-dimensional magnonic crystals that opens the door to new concepts in nano-magnonics.

We used a dynamic FMR approach for detecting the lattice interactions and we primarily focused on harnessing the tunability of these interactions for nano-magnonics. However, we note that the choice of dissimilar materials for different sublattice sites could also be exploited for controlling the order state of artificial spin ice \cite{Mengotti2008,Anghinolfi2015,Farhan2019}. %sTherefore, the presented approach to use different materials in the sublattices could not only be used to control the dynamics, but also to control the state of the ASI.

\section{Supporting Information}

The Supporting Information contains information about the sample fabrication, VNA-FMR measurement technique, micromagnetic simulations, detailed ferromagnetic resonance equations, and additional data for reference films and single-component ASI.

\begin{acknowledgement}

%Please use ``The authors thank \ldots'' rather than ``The
%authors would like to thank \ldots''.

This work was supported by the U.S. Department of Energy, Office of Basic Energy Sciences, Division of Materials Sciences and Engineering under Award DE-SC0020308.

\end{acknowledgement}

\
%%%%%%%%%%%%%%%%%%%%%%%%%%%%%%%%%%%%%%%%%%%%%%%%%%%%%%%%%%%%%%%%%%%%%
%% The same is true for Supporting Information, which should use the
%% suppinfo environment.
%%%%%%%%%%%%%%%%%%%%%%%%%%%%%%%%%%%%%%%%%%%%%%%%%%%%%%%%%%%%%%%%%%%%%
%\begin{suppinfo}
%
%A listing of the contents of each file supplied as Supporting Information
%should be included. For instructions on what should be included in the
%Supporting Information as well as how to prepare this material for
%publications, refer to the journal's Instructions for Authors.
%
%The following files are available free of charge.
%\begin{itemize}
%  \item Filename: brief description
%  \item Filename: brief description
%\end{itemize}
%
%\end{suppinfo}

%%%%%%%%%%%%%%%%%%%%%%%%%%%%%%%%%%%%%%%%%%%%%%%%%%%%%%%%%%%%%%%%%%%%%
%% The appropriate \bibliography command should be placed here.
%% Notice that the class file automatically sets \bibliographystyle
%% and also names the section correctly.
%%%%%%%%%%%%%%%%%%%%%%%%%%%%%%%%%%%%%%%%%%%%%%%%%%%%%%%%%%%%%%%%%%%%%
%\bibliography{bibliography}

\begin{mcitethebibliography}{44}
\providecommand*\natexlab[1]{#1}
\providecommand*\mciteSetBstSublistMode[1]{}
\providecommand*\mciteSetBstMaxWidthForm[2]{}
\providecommand*\mciteBstWouldAddEndPuncttrue
  {\def\EndOfBibitem{\unskip.}}
\providecommand*\mciteBstWouldAddEndPunctfalse
  {\let\EndOfBibitem\relax}
\providecommand*\mciteSetBstMidEndSepPunct[3]{}
\providecommand*\mciteSetBstSublistLabelBeginEnd[3]{}
\providecommand*\EndOfBibitem{}
\mciteSetBstSublistMode{f}
\mciteSetBstMaxWidthForm{subitem}{(\alph{mcitesubitemcount})}
\mciteSetBstSublistLabelBeginEnd
  {\mcitemaxwidthsubitemform\space}
  {\relax}
  {\relax}

\bibitem[Kruglyak \latin{et~al.}(2010)Kruglyak, Demokritov, and
  Grundler]{Kruglyak2010}
Kruglyak,~V.~V.; Demokritov,~S.~O.; Grundler,~D. \emph{J. Phys. D. Appl. Phys.}
  \textbf{2010}, \emph{43}, 264001\relax
\mciteBstWouldAddEndPuncttrue
\mciteSetBstMidEndSepPunct{\mcitedefaultmidpunct}
{\mcitedefaultendpunct}{\mcitedefaultseppunct}\relax
\EndOfBibitem
\bibitem[Chumak \latin{et~al.}(2015)Chumak, Vasyuchka, Serga, and
  Hillebrands]{Chumak_NatPhys_2015}
Chumak,~A.~V.; Vasyuchka,~V.~I.; Serga,~A.~A.; Hillebrands,~B. \emph{Nat.
  Phys.} \textbf{2015}, \emph{11}, 453--461\relax
\mciteBstWouldAddEndPuncttrue
\mciteSetBstMidEndSepPunct{\mcitedefaultmidpunct}
{\mcitedefaultendpunct}{\mcitedefaultseppunct}\relax
\EndOfBibitem
\bibitem[Vogt \latin{et~al.}(2014)Vogt, Fradin, Pearson, Sebastian, Bader,
  Hillebrands, Hoffmann, and Schultheiss]{Vogt2014}
Vogt,~K.; Fradin,~F.; Pearson,~J.; Sebastian,~T.; Bader,~S.; Hillebrands,~B.;
  Hoffmann,~A.; Schultheiss,~H. \emph{Nat. Commun.} \textbf{2014}, \emph{5},
  3727\relax
\mciteBstWouldAddEndPuncttrue
\mciteSetBstMidEndSepPunct{\mcitedefaultmidpunct}
{\mcitedefaultendpunct}{\mcitedefaultseppunct}\relax
\EndOfBibitem
\bibitem[Bonetti and {\AA}kerman(2013)Bonetti, and {\AA}kerman]{Bonetti2013}
Bonetti,~S.; {\AA}kerman,~J. In \emph{Magnonics}; Demokritov,~S.~O.,
  Slavin,~A.~N., Eds.; Topics in Applied Physics; Springer Berlin Heidelberg:
  Berlin, Heidelberg, 2013; Vol. 125; pp 177--187\relax
\mciteBstWouldAddEndPuncttrue
\mciteSetBstMidEndSepPunct{\mcitedefaultmidpunct}
{\mcitedefaultendpunct}{\mcitedefaultseppunct}\relax
\EndOfBibitem
\bibitem[Grollier \latin{et~al.}(2016)Grollier, Querlioz, and
  Stiles]{Grollier2016}
Grollier,~J.; Querlioz,~D.; Stiles,~M.~D. \emph{Proc. IEEE} \textbf{2016},
  \emph{104}, 2024--2039\relax
\mciteBstWouldAddEndPuncttrue
\mciteSetBstMidEndSepPunct{\mcitedefaultmidpunct}
{\mcitedefaultendpunct}{\mcitedefaultseppunct}\relax
\EndOfBibitem
\bibitem[Csaba \latin{et~al.}(2017)Csaba, Papp, and Porod]{Csaba_2017}
Csaba,~G.; Papp,~{\'A}.; Porod,~W. \emph{Physics Letters A} \textbf{2017},
  \emph{381}, 1471--1476\relax
\mciteBstWouldAddEndPuncttrue
\mciteSetBstMidEndSepPunct{\mcitedefaultmidpunct}
{\mcitedefaultendpunct}{\mcitedefaultseppunct}\relax
\EndOfBibitem
\bibitem[Tabuchi \latin{et~al.}(2014)Tabuchi, Ishino, Ishikawa, Yamazaki,
  Usami, and Nakamura]{Tabuchi_2014}
Tabuchi,~Y.; Ishino,~S.; Ishikawa,~T.; Yamazaki,~R.; Usami,~K.; Nakamura,~Y.
  \emph{Phys. Rev. Lett.} \textbf{2014}, \emph{113}, 083603--5\relax
\mciteBstWouldAddEndPuncttrue
\mciteSetBstMidEndSepPunct{\mcitedefaultmidpunct}
{\mcitedefaultendpunct}{\mcitedefaultseppunct}\relax
\EndOfBibitem
\bibitem[Zhang \latin{et~al.}(2016)Zhang, Zou, Jiang, and Tang]{Zhang_2016}
Zhang,~X.; Zou,~C.-L.; Jiang,~L.; Tang,~H.~X. \emph{Science Advances}
  \textbf{2016}, \emph{2}, e1501286\relax
\mciteBstWouldAddEndPuncttrue
\mciteSetBstMidEndSepPunct{\mcitedefaultmidpunct}
{\mcitedefaultendpunct}{\mcitedefaultseppunct}\relax
\EndOfBibitem
\bibitem[Wang \latin{et~al.}(2009)Wang, Zhang, Lim, Ng, Kuok, Jain, and
  Adeyeye]{Wang2009}
Wang,~Z.~K.; Zhang,~V.~L.; Lim,~H.~S.; Ng,~S.~C.; Kuok,~M.~H.; Jain,~S.;
  Adeyeye,~A.~O. \emph{Appl. Phys. Lett.} \textbf{2009}, \emph{94},
  083112\relax
\mciteBstWouldAddEndPuncttrue
\mciteSetBstMidEndSepPunct{\mcitedefaultmidpunct}
{\mcitedefaultendpunct}{\mcitedefaultseppunct}\relax
\EndOfBibitem
\bibitem[Kostylev \latin{et~al.}(2008)Kostylev, Schrader, Stamps, Gubbiotti,
  Carlotti, Adeyeye, Goolaup, and Singh]{Kostylev2008}
Kostylev,~M.; Schrader,~P.; Stamps,~R.~L.; Gubbiotti,~G.; Carlotti,~G.;
  Adeyeye,~A.~O.; Goolaup,~S.; Singh,~N. \emph{Appl. Phys. Lett.}
  \textbf{2008}, \emph{92}, 132504\relax
\mciteBstWouldAddEndPuncttrue
\mciteSetBstMidEndSepPunct{\mcitedefaultmidpunct}
{\mcitedefaultendpunct}{\mcitedefaultseppunct}\relax
\EndOfBibitem
\bibitem[Chumak \latin{et~al.}(2017)Chumak, Serga, and
  Hillebrands]{Chumak_2017}
Chumak,~A.~V.; Serga,~A.~A.; Hillebrands,~B. \emph{J. Phys. D} \textbf{2017},
  \emph{50}, 244001\relax
\mciteBstWouldAddEndPuncttrue
\mciteSetBstMidEndSepPunct{\mcitedefaultmidpunct}
{\mcitedefaultendpunct}{\mcitedefaultseppunct}\relax
\EndOfBibitem
\bibitem[Lenk \latin{et~al.}(2011)Lenk, Ulrichs, Garbs, and
  M{\"u}nzenberg]{Lenk_2011}
Lenk,~B.; Ulrichs,~H.; Garbs,~F.; M{\"u}nzenberg,~M. \emph{Phys. Rep.}
  \textbf{2011}, \emph{507}, 107--136\relax
\mciteBstWouldAddEndPuncttrue
\mciteSetBstMidEndSepPunct{\mcitedefaultmidpunct}
{\mcitedefaultendpunct}{\mcitedefaultseppunct}\relax
\EndOfBibitem
\bibitem[Krawczyk and Grundler(2014)Krawczyk, and Grundler]{Krawczyk_2014}
Krawczyk,~M.; Grundler,~D. \emph{J. Phys.: Condens. Matter} \textbf{2014},
  \emph{26}, 123202--33\relax
\mciteBstWouldAddEndPuncttrue
\mciteSetBstMidEndSepPunct{\mcitedefaultmidpunct}
{\mcitedefaultendpunct}{\mcitedefaultseppunct}\relax
\EndOfBibitem
\bibitem[Lendinez and Jungfleisch(2020)Lendinez, and Jungfleisch]{Lendinez2020}
Lendinez,~S.; Jungfleisch,~M.~B. \emph{J. Phys. Condens. Matter} \textbf{2020},
  \emph{32}, 013001\relax
\mciteBstWouldAddEndPuncttrue
\mciteSetBstMidEndSepPunct{\mcitedefaultmidpunct}
{\mcitedefaultendpunct}{\mcitedefaultseppunct}\relax
\EndOfBibitem
\bibitem[Sklenar \latin{et~al.}(2019)Sklenar, Lendinez, and
  Jungfleisch]{Sklenar2019}
Sklenar,~J.; Lendinez,~S.; Jungfleisch,~M.~B. \emph{Solid State Phys.}
  \textbf{2019}, \emph{70}, 171--235\relax
\mciteBstWouldAddEndPuncttrue
\mciteSetBstMidEndSepPunct{\mcitedefaultmidpunct}
{\mcitedefaultendpunct}{\mcitedefaultseppunct}\relax
\EndOfBibitem
\bibitem[Skj{\ae}rv{\o} \latin{et~al.}(2019)Skj{\ae}rv{\o}, Marrows, Stamps,
  and Heyderman]{Skjaervo_2019}
Skj{\ae}rv{\o},~S.~H.; Marrows,~C.~H.; Stamps,~R.~L.; Heyderman,~L.~J.
  \emph{Nat Rev Phys} \textbf{2019}, \emph{439}, 1--16\relax
\mciteBstWouldAddEndPuncttrue
\mciteSetBstMidEndSepPunct{\mcitedefaultmidpunct}
{\mcitedefaultendpunct}{\mcitedefaultseppunct}\relax
\EndOfBibitem
\bibitem[Gliga \latin{et~al.}(2020)Gliga, Iacocca, and Heinonen]{Gliga2020}
Gliga,~S.; Iacocca,~E.; Heinonen,~O.~G. \emph{APL Materials} \textbf{2020},
  \emph{8}, 040911\relax
\mciteBstWouldAddEndPuncttrue
\mciteSetBstMidEndSepPunct{\mcitedefaultmidpunct}
{\mcitedefaultendpunct}{\mcitedefaultseppunct}\relax
\EndOfBibitem
\bibitem[Dion \latin{et~al.}(2019)Dion, Arroo, Yamanoi, Kimura, Gartside,
  Cohen, Kurebayashi, and Branford]{Dion_2019}
Dion,~T.; Arroo,~D.~M.; Yamanoi,~K.; Kimura,~T.; Gartside,~J.~C.; Cohen,~L.~F.;
  Kurebayashi,~H.; Branford,~W.~R. \emph{Phys. Rev. B} \textbf{2019},
  \emph{100}, 054433\relax
\mciteBstWouldAddEndPuncttrue
\mciteSetBstMidEndSepPunct{\mcitedefaultmidpunct}
{\mcitedefaultendpunct}{\mcitedefaultseppunct}\relax
\EndOfBibitem
\bibitem[Iacocca \latin{et~al.}(2020)Iacocca, Gliga, and
  Heinonen]{Iacocca_2020}
Iacocca,~E.; Gliga,~S.; Heinonen,~O.~G. \emph{Phys. Rev. Applied}
  \textbf{2020}, \emph{13}, 044047\relax
\mciteBstWouldAddEndPuncttrue
\mciteSetBstMidEndSepPunct{\mcitedefaultmidpunct}
{\mcitedefaultendpunct}{\mcitedefaultseppunct}\relax
\EndOfBibitem
\bibitem[Harris \latin{et~al.}(1997)Harris, Bramwell, McMorrow, Zeiske, and
  Godfrey]{Harris1997}
Harris,~M.~J.; Bramwell,~S.~T.; McMorrow,~D.~F.; Zeiske,~T.; Godfrey,~K.~W.
  \emph{Phys. Rev. Lett.} \textbf{1997}, \emph{79}, 2554--2557\relax
\mciteBstWouldAddEndPuncttrue
\mciteSetBstMidEndSepPunct{\mcitedefaultmidpunct}
{\mcitedefaultendpunct}{\mcitedefaultseppunct}\relax
\EndOfBibitem
\bibitem[Bramwell and Gingras(2001)Bramwell, and Gingras]{Bramwell2001}
Bramwell,~S.~T.; Gingras,~M.~J. \emph{Science} \textbf{2001}, \emph{294},
  1495--501\relax
\mciteBstWouldAddEndPuncttrue
\mciteSetBstMidEndSepPunct{\mcitedefaultmidpunct}
{\mcitedefaultendpunct}{\mcitedefaultseppunct}\relax
\EndOfBibitem
\bibitem[Branford \latin{et~al.}(2012)Branford, Ladak, Read, Zeissler, and
  Cohen]{Branford2012}
Branford,~W.~R.; Ladak,~S.; Read,~D.~E.; Zeissler,~K.; Cohen,~L.~F.
  \emph{Science} \textbf{2012}, \emph{335}, 1597--600\relax
\mciteBstWouldAddEndPuncttrue
\mciteSetBstMidEndSepPunct{\mcitedefaultmidpunct}
{\mcitedefaultendpunct}{\mcitedefaultseppunct}\relax
\EndOfBibitem
\bibitem[Gilbert \latin{et~al.}(2016)Gilbert, Nisoli, and
  Schiffer]{Gilbert2016}
Gilbert,~I.; Nisoli,~C.; Schiffer,~P. \emph{Phys. Today} \textbf{2016},
  \emph{69}, 54--59\relax
\mciteBstWouldAddEndPuncttrue
\mciteSetBstMidEndSepPunct{\mcitedefaultmidpunct}
{\mcitedefaultendpunct}{\mcitedefaultseppunct}\relax
\EndOfBibitem
\bibitem[Sklenar \latin{et~al.}(2013)Sklenar, Bhat, DeLong, and
  Ketterson]{Sklenar2013}
Sklenar,~J.; Bhat,~V.~S.; DeLong,~L.~E.; Ketterson,~J.~B. \emph{J. Appl. Phys.}
  \textbf{2013}, \emph{113}, 17B530\relax
\mciteBstWouldAddEndPuncttrue
\mciteSetBstMidEndSepPunct{\mcitedefaultmidpunct}
{\mcitedefaultendpunct}{\mcitedefaultseppunct}\relax
\EndOfBibitem
\bibitem[Gliga \latin{et~al.}(2013)Gliga, K{\'{a}}kay, Hertel, and
  Heinonen]{Gliga2013}
Gliga,~S.; K{\'{a}}kay,~A.; Hertel,~R.; Heinonen,~O.~G. \emph{Phys. Rev. Lett.}
  \textbf{2013}, \emph{110}, 117205\relax
\mciteBstWouldAddEndPuncttrue
\mciteSetBstMidEndSepPunct{\mcitedefaultmidpunct}
{\mcitedefaultendpunct}{\mcitedefaultseppunct}\relax
\EndOfBibitem
\bibitem[Jungfleisch \latin{et~al.}(2016)Jungfleisch, Zhang, Iacocca, Sklenar,
  Ding, Jiang, Zhang, Pearson, Novosad, Ketterson, Heinonen, and
  Hoffmann]{Jungfleisch2016}
Jungfleisch,~M.~B.; Zhang,~W.; Iacocca,~E.; Sklenar,~J.; Ding,~J.; Jiang,~W.;
  Zhang,~S.; Pearson,~J.~E.; Novosad,~V.; Ketterson,~J.~B.; Heinonen,~O.;
  Hoffmann,~A. \emph{Phys. Rev. B} \textbf{2016}, \emph{93}, 100401\relax
\mciteBstWouldAddEndPuncttrue
\mciteSetBstMidEndSepPunct{\mcitedefaultmidpunct}
{\mcitedefaultendpunct}{\mcitedefaultseppunct}\relax
\EndOfBibitem
\bibitem[Zhou \latin{et~al.}(2016)Zhou, Chua, Singh, and Adeyeye]{Zhou2016}
Zhou,~X.; Chua,~G.-L.; Singh,~N.; Adeyeye,~A.~O. \emph{Adv. Funct. Mater.}
  \textbf{2016}, \emph{26}, 1437--1444\relax
\mciteBstWouldAddEndPuncttrue
\mciteSetBstMidEndSepPunct{\mcitedefaultmidpunct}
{\mcitedefaultendpunct}{\mcitedefaultseppunct}\relax
\EndOfBibitem
\bibitem[Iacocca \latin{et~al.}(2016)Iacocca, Gliga, Stamps, and
  Heinonen]{Iacocca_PRB2016}
Iacocca,~E.; Gliga,~S.; Stamps,~R.~L.; Heinonen,~O. \emph{Phys. Rev. B}
  \textbf{2016}, \emph{93}, 134420\relax
\mciteBstWouldAddEndPuncttrue
\mciteSetBstMidEndSepPunct{\mcitedefaultmidpunct}
{\mcitedefaultendpunct}{\mcitedefaultseppunct}\relax
\EndOfBibitem
\bibitem[Gubbiotti \latin{et~al.}(2018)Gubbiotti, Zhou, Haghshenasfard, Cottam,
  and Adeyeye]{Gubbiotti2018}
Gubbiotti,~G.; Zhou,~X.; Haghshenasfard,~Z.; Cottam,~M.~G.; Adeyeye,~A.~O.
  \emph{Phys. Rev. B} \textbf{2018}, \emph{97}, 134428\relax
\mciteBstWouldAddEndPuncttrue
\mciteSetBstMidEndSepPunct{\mcitedefaultmidpunct}
{\mcitedefaultendpunct}{\mcitedefaultseppunct}\relax
\EndOfBibitem
\bibitem[{\"{O}}stman \latin{et~al.}(2018){\"{O}}stman, Stopfel, Chioar,
  Arnalds, Stein, Kapaklis, and Hj{\"{o}}rvarsson]{Ostman2018}
{\"{O}}stman,~E.; Stopfel,~H.; Chioar,~I.-A.; Arnalds,~U.~B.; Stein,~A.;
  Kapaklis,~V.; Hj{\"{o}}rvarsson,~B. \emph{Nat. Phys.} \textbf{2018},
  \emph{14}, 375--379\relax
\mciteBstWouldAddEndPuncttrue
\mciteSetBstMidEndSepPunct{\mcitedefaultmidpunct}
{\mcitedefaultendpunct}{\mcitedefaultseppunct}\relax
\EndOfBibitem
\bibitem[Drisko \latin{et~al.}(2017)Drisko, Marsh, and Cumings]{Drisko2017}
Drisko,~J.; Marsh,~T.; Cumings,~J. \emph{Nat. Commun.} \textbf{2017}, \emph{8},
  14009\relax
\mciteBstWouldAddEndPuncttrue
\mciteSetBstMidEndSepPunct{\mcitedefaultmidpunct}
{\mcitedefaultendpunct}{\mcitedefaultseppunct}\relax
\EndOfBibitem
\bibitem[Schanilec \latin{et~al.}(2019)Schanilec, Perrin, Denmat, Canals, and
  Rougemaille]{Schanilec2019}
Schanilec,~V.; Perrin,~Y.; Denmat,~S.~L.; Canals,~B.; Rougemaille,~N.
  \emph{arXiv} \textbf{2019}, 1902.00452\relax
\mciteBstWouldAddEndPuncttrue
\mciteSetBstMidEndSepPunct{\mcitedefaultmidpunct}
{\mcitedefaultendpunct}{\mcitedefaultseppunct}\relax
\EndOfBibitem
\bibitem[Farhan \latin{et~al.}(2019)Farhan, Saccone, Petersen, Dhuey,
  Chopdekar, Huang, Kent, Chen, Alava, Lippert, Scholl, and van
  Dijken]{Farhan2019}
Farhan,~A.; Saccone,~M.; Petersen,~C.~F.; Dhuey,~S.; Chopdekar,~R.~V.;
  Huang,~Y.-L.; Kent,~N.; Chen,~Z.; Alava,~M.~J.; Lippert,~T.; Scholl,~A.; van
  Dijken,~S. \emph{Sci. Adv.} \textbf{2019}, \emph{5}, eaav6380\relax
\mciteBstWouldAddEndPuncttrue
\mciteSetBstMidEndSepPunct{\mcitedefaultmidpunct}
{\mcitedefaultendpunct}{\mcitedefaultseppunct}\relax
\EndOfBibitem
\bibitem[Vansteenkiste \latin{et~al.}(2014)Vansteenkiste, Leliaert, Dvornik,
  Helsen, Garcia-Sanchez, and {Van Waeyenberge}]{Vansteenkiste2014}
Vansteenkiste,~A.; Leliaert,~J.; Dvornik,~M.; Helsen,~M.; Garcia-Sanchez,~F.;
  {Van Waeyenberge},~B. \emph{AIP Adv.} \textbf{2014}, \emph{4}, 107133\relax
\mciteBstWouldAddEndPuncttrue
\mciteSetBstMidEndSepPunct{\mcitedefaultmidpunct}
{\mcitedefaultendpunct}{\mcitedefaultseppunct}\relax
\EndOfBibitem
\bibitem[Kalarickal \latin{et~al.}(2006)Kalarickal, Krivosik, Wu, Patton,
  Schneider, Kabos, Silva, and Nibarger]{Kalarickal2006}
Kalarickal,~S.~S.; Krivosik,~P.; Wu,~M.; Patton,~C.~E.; Schneider,~M.~L.;
  Kabos,~P.; Silva,~T.~J.; Nibarger,~J.~P. \emph{Journal of Applied Physics}
  \textbf{2006}, \emph{99}, 093909\relax
\mciteBstWouldAddEndPuncttrue
\mciteSetBstMidEndSepPunct{\mcitedefaultmidpunct}
{\mcitedefaultendpunct}{\mcitedefaultseppunct}\relax
\EndOfBibitem
\bibitem[Suhl(1955)]{Suhl1955}
Suhl,~H. \emph{Phys. Rev.} \textbf{1955}, \emph{97}, 555--557\relax
\mciteBstWouldAddEndPuncttrue
\mciteSetBstMidEndSepPunct{\mcitedefaultmidpunct}
{\mcitedefaultendpunct}{\mcitedefaultseppunct}\relax
\EndOfBibitem
\bibitem[Montoncello \latin{et~al.}(2018)Montoncello, Giovannini, Bang,
  Ketterson, Jungfleisch, Hoffmann, Farmer, and De~Long]{Montoncello_2018}
Montoncello,~F.; Giovannini,~L.; Bang,~W.; Ketterson,~J.~B.;
  Jungfleisch,~M.~B.; Hoffmann,~A.; Farmer,~B.~W.; De~Long,~L.~E. \emph{Phys.
  Rev. B} \textbf{2018}, \emph{97}, 014421\relax
\mciteBstWouldAddEndPuncttrue
\mciteSetBstMidEndSepPunct{\mcitedefaultmidpunct}
{\mcitedefaultendpunct}{\mcitedefaultseppunct}\relax
\EndOfBibitem
\bibitem[Bang \latin{et~al.}(2020)Bang, Sturm, Silvani, Kaffash, Hoffmann,
  Ketterson, Montoncello, and Jungfleisch]{Bang_2020}
Bang,~W.; Sturm,~J.; Silvani,~R.; Kaffash,~M.~T.; Hoffmann,~A.;
  Ketterson,~J.~B.; Montoncello,~F.; Jungfleisch,~M.~B. \emph{Phys. Rev.
  Applied} \textbf{2020}, \emph{14}, 014079\relax
\mciteBstWouldAddEndPuncttrue
\mciteSetBstMidEndSepPunct{\mcitedefaultmidpunct}
{\mcitedefaultendpunct}{\mcitedefaultseppunct}\relax
\EndOfBibitem
\bibitem[Shiota \latin{et~al.}(2020)Shiota, Taniguchi, Ishibashi, Moriyama, and
  Ono]{Ono}
Shiota,~Y.; Taniguchi,~T.; Ishibashi,~M.; Moriyama,~T.; Ono,~T. \emph{Phys.
  Rev. Lett.} \textbf{2020}, \emph{125}, 017203\relax
\mciteBstWouldAddEndPuncttrue
\mciteSetBstMidEndSepPunct{\mcitedefaultmidpunct}
{\mcitedefaultendpunct}{\mcitedefaultseppunct}\relax
\EndOfBibitem
\bibitem[Li \latin{et~al.}(2020)Li, Cao, Amin, Zhang, Gibbons, Sklenar,
  Pearson, Haney, Stiles, Bailey, Novosad, Hoffmann, and Zhang]{Li_2020}
Li,~Y.; Cao,~W.; Amin,~V.~P.; Zhang,~Z.; Gibbons,~J.; Sklenar,~J.; Pearson,~J.;
  Haney,~P.~M.; Stiles,~M.~D.; Bailey,~W.~E.; Novosad,~V.; Hoffmann,~A.;
  Zhang,~W. \emph{Phys. Rev. Lett.} \textbf{2020}, \emph{124}, 117202\relax
\mciteBstWouldAddEndPuncttrue
\mciteSetBstMidEndSepPunct{\mcitedefaultmidpunct}
{\mcitedefaultendpunct}{\mcitedefaultseppunct}\relax
\EndOfBibitem
\bibitem[MacNeill \latin{et~al.}(2019)MacNeill, Hou, Klein, Zhang,
  Jarillo-Herrero, and Liu]{Liu_2019}
MacNeill,~D.; Hou,~J.~T.; Klein,~D.~R.; Zhang,~P.; Jarillo-Herrero,~P.; Liu,~L.
  \emph{Phys. Rev. Lett.} \textbf{2019}, \emph{123}, 047204\relax
\mciteBstWouldAddEndPuncttrue
\mciteSetBstMidEndSepPunct{\mcitedefaultmidpunct}
{\mcitedefaultendpunct}{\mcitedefaultseppunct}\relax
\EndOfBibitem
\bibitem[Mengotti \latin{et~al.}(2008)Mengotti, Heyderman, {Fraile
  Rodr{\'{i}}guez}, Bisig, {Le Guyader}, Nolting, and Braun]{Mengotti2008}
Mengotti,~E.; Heyderman,~L.~J.; {Fraile Rodr{\'{i}}guez},~A.; Bisig,~A.; {Le
  Guyader},~L.; Nolting,~F.; Braun,~H.~B. \emph{Phys. Rev. B} \textbf{2008},
  \emph{78}, 144402\relax
\mciteBstWouldAddEndPuncttrue
\mciteSetBstMidEndSepPunct{\mcitedefaultmidpunct}
{\mcitedefaultendpunct}{\mcitedefaultseppunct}\relax
\EndOfBibitem
\bibitem[Anghinolfi \latin{et~al.}(2015)Anghinolfi, Luetkens, Perron, Flokstra,
  Sendetskyi, Suter, Prokscha, Derlet, Lee, and Heyderman]{Anghinolfi2015}
Anghinolfi,~L.; Luetkens,~H.; Perron,~J.; Flokstra,~M.~G.; Sendetskyi,~O.;
  Suter,~A.; Prokscha,~T.; Derlet,~P.~M.; Lee,~S.~L.; Heyderman,~L.~J.
  \emph{Nat. Commun.} \textbf{2015}, \emph{6}, 8278\relax
\mciteBstWouldAddEndPuncttrue
\mciteSetBstMidEndSepPunct{\mcitedefaultmidpunct}
{\mcitedefaultendpunct}{\mcitedefaultseppunct}\relax
\EndOfBibitem
\end{mcitethebibliography}

\providecommand{\latin}[1]{#1}
\makeatletter
\providecommand{\doi}
  {\begingroup\let\do\@makeother\dospecials
  \catcode`\{=1 \catcode`\}=2 \doi@aux}
\providecommand{\doi@aux}[1]{\endgroup\texttt{#1}}
\makeatother
\providecommand*\mcitethebibliography{\thebibliography}
\csname @ifundefined\endcsname{endmcitethebibliography}
  {\let\endmcitethebibliography\endthebibliography}{}

\end{document}